\documentclass[aps,pre,twocolumn,showpacs,amssymb]{revtex4}

\usepackage{graphicx,bm,times}

\begin{document}

\title{Processed Splitting Algorithms for Rigid-Body
Molecular Dynamics Simulations}

\author{Igor P. Omelyan}

\affiliation{Institute for Condensed Matter Physics,
1 Svientsitskii Street, UA-79011 Lviv, Ukraine}

\affiliation{{Institute for Theoretical Physics, Linz University,
A-4040 Linz, Austria}}

\date{\today}

\begin{abstract}

A new approach for integration of motion in many-body systems of
interacting polyatomic molecules is proposed. It is based on
splitting time propagation of pseudo-variables in a modified phase
space, while the real translational and orientational coordinates
are decoded by processing transformations. This allows to overcome
the barrier on the order of precision of the integration at a given
number of force-torque evaluations per time step. Testing in
dynamics of water versus previous methods shows that the obtained
algorithms significantly improve the accuracy of the simulations
without extra computational costs.

\end{abstract}


\maketitle

\section{Introduction}

Systems of rigid bodies are widely used to model various phenomena
on a broad range of length scales: from the microscopic dynamics of
molecules in gases and liquids \cite{Allen,Rapaport}, mesoscopic
behavior of polymers and other complex collections in chemical and
biological physics \cite{Frenkel,Essiz} to macroscopic movement of
astrophysical objects in celestial mechanics \cite{Celledoni,Chin}.
A lot of approaches, including the traditional Runge-Kutta and
predictor-corrector schemes \cite{Allen} as well as more recent
splitting techniques \cite{Reichs,Kol,Dullw,Matubayasi,Miller,%
Kamberaj}, have been devised over the years to integrate the
rigid-body equations of motion.

Now it is well established that the most adequate integration can be
done by splitting the time propagator into analytically solvable
parts \cite{Omfcpcn,Omepre,Omejcp}. For Hamiltonian systems this
provides the preservation of such essential properties as
conservation of volume in phase space and time reversibility. As a
result, the splitting algorithms exhibit remarkable stability and
thus are ideal for long-duration molecular dynamics (MD)
simulations. In addition, these algorithms can be symplectic, i.e.
can exactly conserve the total energy associated with a nearby
Hamiltonian.

The splitting approach however has a limitation on the order $K$ of
precision at each given number $n$ of force-torque evaluations per
time step. Note that these evaluations present the most
time-consuming part of the propagation. For this reason, the
rigid-body motion in MD simulations is integrated mainly by the
simplest ($K=2$) Verlet-type algorithms \cite{Kol,Dullw,%
Matubayasi,Miller,Kamberaj} with $n=1$. The optimized algorithms
\cite{Omfcpcn,Omepre,Omejcp} at $n=2$ can outperform Verlet schemes.
But such an optimization does not rise the order of precision and
for $K=2$ only modest accuracy can be reached. Higher-order ($K=4$)
splitting schemes (note that $K$ should be even to ensure time
reversibility) can be derived beginning from $n=3$ \cite{Omepre,%
Omejcp}. The grown computational costs at $n=3$ and $K=4$ can be
compensated by the increased precision when adding gradient-like
terms to the splitting propagator \cite{Omejcp}.

Meanwhile it has been found that the order $K$ of precision can be
risen by carrying out supplementary (so-called processed)
decompositions apart from the basic (kernel) splitting
\cite{Blanes}. For $K=4$, each minimal kernel and processor leads to
one force and one force-gradient evaluations. This yields an
effective number $n=2(1+\nu)$, where $\nu$ is the relative cost
spent on the gradient evaluation with respect to that on the force
calculation. This number can be decreased twice to $n=1+\nu$ by
constructing cheap approximate processors \cite{Skeel,Marcos,Casas}.
Taking into account that the evaluation of one force gradient is
more expensive at least in a factor of $\nu=2$ than the calculation
of one force \cite{Omfcpcn,Omepre,Omejcp} gives that $n \ge 3$.
However, the gradient evaluation may present a difficulty for
systems with long-range (e.g. Coulomb) interactions, where the
factor $\nu$ can be too large \cite{Omejcp} because of the necessity
to calculate cumbersome tail (Ewald-summated) contributions. Note
also that the processed algorithms of Refs.~\cite{Skeel,Marcos} were
obtained exclusively for pure translational motion and they are not
suitable for rigid-body dynamics. The processing methods introduced
in Refs.~\cite{Blanes,Casas} for solving ordinary differential
equations are more general but need an adaptation to be exploited in
the case of rotational motion. In particular, contrary to free
translational dynamics, the propagator of free rotational motion
cannot be handled at once and requires additional splitting into
analytically integrable parts \cite{Omejcp} or involving special
functions \cite{Ramses}.

Up to now, no processing schemes were designed and applied to MD
simulations of interacting rigid bodies. The rotational motion is
much more complicated than translational displacements and thus
demands a separate investigation. Moreover, a fundamental
theoretical problem on the possibility to overcome the barrier $n=3$
for the fourth-order integration still remains open. This overcoming
is important from the practical point of view as well, because
smaller values of $n$ could noticeably speed up the calculations in
view of the restricted capabilities of even supercomputers.

In the proposing paper we develop the processing formalism in the
explicit presence of translational and orientational degrees of
freedom. We show that using a proper transformation of phase
coordinates allows to lower the fourth-order barrier to the value
$n=2$ with no gradient evaluations. It is proven also that in a
specific case of quasi-fourth-order integration the number of
force-gradient evaluations per step can be reduced to $n=1$ at all.

The paper is organized as follows. The new processed algorithms are
consistently derived in Sec. II. Their applications to rigid-body MD
simulations and comparison with integrators known previously are
presented in Sec. III. Concluding remarks are highlighted in Sec.
IV.

\section{Theory}

Let us consider a classical system of $N$ interacting rigid
polyatomic molecules. The dynamical state of such a system in the
laboratory frame is determined by the position ${\bf r}_i$ of the
center of mass $m$ of the $i$th molecule, its attitude matrix ${\bf
S}_i$ as well as the translational ${\bf p}_i$ and angular ${\bf
q}_i$ momenta. The equations of motion can be written in the
following compact form $d \bm{\rho}/d t = L \bm{\rho}(t)$. Here
$\bm{\rho} = \{{\bf r}_1, {\bf p}_1, {\bf S}_1, {\bf q}_1; \ldots;
{\bf r}_N, {\bf p}_N, {\bf S}_N, {\bf q}_N \} \equiv \{{\bf r}, {\bf
p}, {\bf S}, {\bf q}\}$ is the set of phase variables,
\begin{eqnarray}
L = \sum_{i=1}^N \!&\bigg[&\!
\frac{{\bf p}_i}{m} \bm{\cdot} \frac{\partial}{\partial {\bf r}_i}
+ {\bf W}({\bf J}^{-1} {\bf S}_i {\bf q}_i) {\bf S}_i \bm{\cdot}
\frac{\partial}{\partial {\bf S}_i}
\nonumber \\ [-7pt] \\ [-7pt]
&+& {\bf f}_i({\bf r},{\bf S}) \bm{\cdot} \frac{\partial}{\partial
{\bf p}_i} + {\bf g}_i({\bf r},{\bf S}) \bm{\cdot} \frac{\partial}
{\partial {\bf q}_i} \bigg] \ \ \ \ \
\nonumber
\end{eqnarray}
denotes the Liouville operator, ${\bf f}_i$ and ${\bf g}_i$ are the
force and torque, respectively, acting on the molecule due to atomic
interactions,
$$
{\bf W}({\bf \Omega})
=\left(
\begin{array}{ccc}
0 & \Omega_Z & -\Omega_Y \\
-\Omega_Z & 0 & \Omega_X \\
\Omega_Y & -\Omega_X & 0
\end{array}
\right)
$$
is the skewsymmetric matrix related to the principal components
$(\Omega_X, \Omega_Y, \Omega_Z)$ of the angular velocity ${\bf
\Omega} = {\bf J}^{-1} {\bf S} {\bf q}$ with ${\bf J}={\rm
diag}(J_X,J_Y,J_Z)$ being the matrix of moments of inertia. If an
initial configuration $\bm{\rho}(0)$ is specified, the unique
solution to the equations of motion can formally be cast for any
time $t$ as $\bm{\rho}(t) = [{\exp}(L h)]^k \bm{\rho}(0)$, where
$h=t/k$ is the size of the time step and $k$ denotes the total
number of steps.

In the standard splitting approach \cite{Omfcpcn,Omepre,Omejcp}, the
Liouville operator $L=A+B$ is decomposed into its kinetic $A=m^{-1}
{\bf p} {\bm \cdot} \partial/\partial {\bf r} + {\bf W}({\bf
\Omega}) {\bf S} {\bm \cdot} \partial/\partial {\bf S}$ and
potential $B={\bf f}({\bf r},{\bf S}) {\bm \cdot} \partial/\partial
{\bf p} + {\bf g}({\bf r},{\bf S}) {\bm \cdot} \partial/\partial
{\bf q}$ parts (we will omit the subscript $i$ for the sake of
simplicity). Then the one-step time propagator ${\rm e}^{L h}$ can
be factorized as ${\rm e}^{(A+B) h + {\mathcal O}(h^{K+1})} =
\prod_{\mu=1}^{n+1} {\rm e}^{B b_\mu h} {\rm e}^{A a_\mu h} \equiv
\Phi_K(h)$, where $n \ge 1$ and $\{a_\mu,b_\mu\}$ are chosen in such
a way to provide the highest possible order $K$ of precision, and
${\mathcal O}(h^{K+1})$ denotes the local error. For instance, the
second-order ($K=2$) Verlet algorithm is obtained at $n=1$ by ${\rm
e}^{(A+B) h + {\mathcal O}(h^3)} = {\rm e}^{B \frac{h}{2}} {\rm
e}^{A h} {\rm e}^{B \frac{h}{2}} \equiv \Phi_2(h)$. Note that the
decomposition constants $a_\mu$ and $b_\mu$ should enter
symmetrically into the factorization to ensure its time
reversibility. This reduces the total number of independent
constants from $2(n+1)$ to $n+1$. In turn the symmetry provides
automatic cancellation of all even-order terms in ${\mathcal
O}(h^{K+1})$, leading to evenness of $K$. For even orders $K \ge 2$,
the local error function has the form ${\mathcal O}(h^{K+1})=c_1
[A,[A,B]] h^3+c_2 [B,[A,B]] h^3+{\mathcal O}(h^{K+3})$, where $[ \ ,
\ ]$ designates the commutator operation and the coefficients $c_1$
and $c_2$ depend on $\{a_\mu,b_\mu\}$. At $K=2$, the two conditions
$\sum_\mu a_\mu = \sum_\mu b_\mu = 1$ should be satisfied to exclude
the zeroth-order term from ${\mathcal O}(h^{K+1})$. In order to
increase the precision to $K=4$ we should satisfy the two additional
conditions $c_1(\{a_\mu,b_\mu\})=c_2(\{a_\mu,b_\mu\})=0$. This can
be provided by increasing the number $n+1$ of independent constants
at least to the number of the order conditions, i.e, to $4$. We see
thus that fourth-order ($K=4$) schemes can be constructed only
beginning from $n=3$ and this number cannot be lowered within the
standard splitting method. At $n=3$, the fourth-order ($K=4$)
factorization can be presented as the concatenation
$\Phi_4(h)=\Phi_2(\chi h)\Phi_2(1-2\chi) h) \Phi_2(\chi h)+{\mathcal
O}(h^5)$ of three Verlet signatures, where $\chi=1/(2-\sqrt{2})$.

For arbitrary times $t$, the solution to the equations of motion can
be evaluated by consecutively applying $k$ times the one-step
splitting propagation $\Phi_K(h)$. This yields $\bm{\rho}(t) =
[\Phi_K(h)]^k \bm{\rho}(0)+ {\mathcal O}(h^K)$, where ${\mathcal
O}(h^{K}) \sim k {\mathcal O}(h^{K+1})$ is the global error due to
the accumulation of the local one after $k=t/h \gg 1$ steps. The
action of the exponential operators ${\rm e}^{A \tau}$ and ${\rm
e}^{B \tau}$ on a phase space point $\bm{\rho}$ is given
analytically by
\begin{eqnarray}
{\rm e}^{A \tau} \big\{{\bf r}, {\bf p}, {\bf S}, {\bf q}\big\} &=&
\big\{ {\bf r} + m^{-1} {\bf p} \tau, {\bf p}, {\bf \Xi}({\bf q},\tau)
{\bf S}, {\bf q} \big\} ,
\nonumber \\ [-6pt] \\ [-4pt]
{\rm e}^{B \tau} \big\{ {\bf r}, {\bf p}, {\bf S}, {\bf q}\big\}
&=& \big\{ {\bf r}, {\bf p} + {\bf f}({\bf r},{\bf S}) \tau, {\bf S},
{\bf q} + {\bf g}({\bf r},{\bf S}) \tau \big\} ,
\ \ \ \ \nonumber
\end{eqnarray}
were the shift of ${\bf r}$ corresponds to free translational motion
(at constant ${\bf p}$), while the changes in ${\bf p}$ and ${\bf
q}$ relate to motion in instantaneous force-torque fields
\cite{Omejcp}. The matrix ${\bf \Xi}({\bf q},\tau)$ exactly
propagates ${\bf S}$ over time $\tau$ according to the free
rotational dynamics (${\bf q}$ remains constant) $d {\bf S}/d t =
{\bf W}({\bf J}^{-1} {\bf S} {\bf q}) {\bf S}$. Expressions for
${\bf \Xi}({\bf q},\tau)$ in terms of efficient routines for
elliptic and theta functions are reported in Ref.~\cite{Ramses}.
Alternatively, ${\bf \Xi}({\bf q},\tau)$ can be replaced by its
second- or fourth-order counterparts ${\bf \Xi}_2(\tau)={\bf
\Psi}_X(\frac{\tau} {2}) {\bf \Psi}_Y(\frac{\tau}{2}){\bf
\Psi}_Z(\tau) {\bf \Psi}_Y (\frac{\tau}{2}){\bf \Psi}_X(\frac
{\tau}{2})$ and ${\bf \Xi}_4(\tau)={\bf \Xi}_2(\chi \tau) {\bf
\Xi}_2((1-2\chi) \tau) {\bf \Xi}_2(\chi \tau)$, where ${\bf
\Psi}_\zeta(\tau)= \exp[{\bf W}(\Omega_\zeta) \tau] \equiv {\bf
\Theta}(\Omega_\zeta,\tau)$ is the matrix representing rotation on
angle $\Omega_\zeta \tau$ around axis $\zeta$ at constant component
$\Omega_\zeta$ of ${\bf \Omega} = {\bf J}^{-1} {\bf S} {\bf q}$ (see
Eq.~(19) of Ref.~\cite{Omejcp} for ${\bf \Theta}(\Omega_\zeta,
\tau)$). Note that each force-torque recalculation in ${\rm e}^{B
\tau}$ requires $\propto N^2$ operations that is the most
time-taking part of the splitting propagation, while the costs for
handling ${\rm e}^{A \tau}$ are negligible (proportional to $N$).
The total number of force-torque recalculations per step in $\Phi_K$
is equal to $n$.

The commutators $[A,[A,B]]$ and $[B,[A,B]]$ which appear in the
local error function ${\mathcal O}(h^{K+1})$ can be calculated
explicitly using the expressions for operators $A$ and $B$. Then,
in the case of the Verlet algorithm ($K=2$) we find $c_1=1/12=2
c_2$ and ${\mathcal O}(h^3) = -( 2 m^{-1} {\bf \dot f} \bm{\cdot}
\partial/\partial {\bf r} - {\bf \ddot f} \bm{\cdot} \partial
/\partial {\bf p}) h^3/12+{\mathcal O}(h^5)$, where at the moment
the orientational degrees of freedom were frozen to simplify
notation. Transferring now the corresponding parts of ${\mathcal
O}(h^3)$ from ${\rm e}^{ L h+ {\mathcal O}(h^3)}$ to the right under
the exponentials ${\rm e}^{A h}$ and ${\rm e}^{B \frac{h}{2}}$ one
obtains ${\rm e}^{L h}={\rm e}^{{\mathcal B} \frac{h}{2}} {\rm
e}^{{\mathcal A} h} {\rm e}^{{\mathcal B} \frac{h}{2}}+{\mathcal
O}(h^5)$, where ${\mathcal A}=A+m^{-1} {\bf \dot f} \bm{\cdot}
\partial/\partial {\bf r} h^2/6$ and $\mathcal{B}=B-{\bf \ddot f}
\bm{\cdot} \partial/\partial {\bf p} h^2/12$ are the modified
counterparts of $A$ and $B$. Thus, the order of the Verlet signature
can increase from $K=2$ to $K=4$ when the decomposition is performed
for the nearby Liouvillian ${\mathcal L}={\mathcal A}+{\mathcal B}=
L(1+m^{-1} {\bf f} \bm{\cdot} \partial/\partial {\bf r} h^2/6 - {\bf
\dot f} \bm{\cdot} \partial/\partial {\bf p} h^2/12)$, where the
equalities ${\bf \dot f}= d {\bf f}/d t = L {\bf f}$ and ${\bf \ddot
f}= L {\bf \dot f}$ for the time derivatives of ${\bf f}$ have been
applied. Note however that the nearby exponentials ${\rm e}^{{
\mathcal A} \tau}$ and ${\rm e}^{{\mathcal B} \tau}$ cannot be
handled analytically in ${\bm \rho}$-space (unlike ${\rm e}^{A
\tau}$ and ${\rm e}^{B \tau}$, see Eq.~(2)), because of the
existence of complicated functions ${\bf \dot f} \equiv {\bf \dot
f}(\bm \rho)$ and ${\bf \ddot f} \equiv {\bf \ddot f}(\bm \rho)$
which contrary to the force field ${\bf f}({\bf r},{\bf S})$ depend
not only on the positions $({\bf r},{\bf S})$ but on the momenta
$({\bf p}, {\bf q})$ as well.

The main idea of our approach consists in finding such a processing
transformation $\bm{\tilde \rho}=\mathfrak{T} {\bm \rho}$ from the
phase space point ${\bm \rho}$ to a new set $\bm{ \tilde \rho}$ of
variables to make the action of the nearby exponentials analytically
calculable. Taking into account the explicit structure for the nearby
Liouvillian ${\mathcal L}$, the general form of the desired
transformation reads $\mathfrak{T}=({\bf r}+\alpha m^{-1} {\bf f}
h^2) \partial/\partial {\bf r} + ({\bf p} + \beta {\bf \dot f} h^2)
\partial/\partial {\bf p}+{\mathcal O}(h^4) \equiv \mathfrak{T}_{
\alpha,\beta}$, where $\alpha$ and $\beta$ are some coefficients
which will be defined below. It can be verified readily that in the
new variables, the equations of motion become $d \bm{\tilde \rho}/d
t = \tilde L \bm{\tilde \rho}$, where $\tilde L= \tilde A + \tilde
B$ is the corresponding Liouville operator with $\tilde A = m^{-1}
[{\bf \tilde p} + (\alpha-\beta) {\bf \dot f}({\bf \tilde r}) h^2]
{\bm \cdot} \partial/\partial {\bf \tilde r}$ and $\tilde B = [{\bf
f}({\bf \tilde r} - \alpha m^{-1} {\bf f}({\bf \tilde r}) h^2) +
\beta {\bf \ddot f}({\bf \tilde r}) h^2] {\bm \cdot} \partial/
\partial {\bf \tilde p}$. Then for the nearby counterparts of
$\tilde A$ and $\tilde B$ one finds $\tilde {\mathcal A}=\tilde
A+m^{-1} {\bf \dot f}({\bf \tilde r}) \bm{\cdot} \partial/
\partial {\bf \tilde r} h^2/6$ and $\tilde {\mathcal B}=\tilde B
-{\bf \ddot f}({\bf \tilde r}) \bm{\cdot} \partial/\partial {\bf
\tilde p} h^2/12$. We see that the terms with ${\bf \dot f}$ and
${\bf \ddot f}$ can be killed in $\tilde {\mathcal A}$ and $\tilde
{\mathcal B}$ by putting $(\alpha - \beta) = - 1/6$ and $\beta=
1/12$, i.e. $\alpha=-1/12$. The orientational degrees of freedom
can be included in a similar manner leading to the total processing
transformation $\mathfrak{T}_{\alpha,\beta}=({\bf r}+\alpha m^{-1}
{\bf f} h^2) \partial/\partial {\bf r} + ({\bf p} + \beta {\bf
\dot f} h^2) \partial/\partial {\bf p} + {\bf \Theta}\big({\bf
J}^{-1}{\bf S}{\bf g}({\bf r},{\bf S}), \alpha h^2 \big) {\bf S}
\partial/\partial {\bf S} + ({\bf q} + \beta {\bf \dot g} h^2)
\partial/\partial {\bf q} +{\mathcal O}(h^4)$ and the nearby
operators $\tilde {\mathcal A} = m^{-1} {\bf \tilde p} {\bm \cdot}
\partial/\partial {\bf \tilde r} + {\bf W}({\bf J}^{-1} {\bf \tilde
S} {\bf \tilde q}) {\bf \tilde S} {\bm \cdot} \partial/\partial {\bf
\tilde S}$ and $\tilde {\mathcal B} = {\bf f}({\bf \tilde r}_\gamma,
{\bf \tilde S}_\gamma) {\bm \cdot} \partial/\partial {\bf \tilde p}
+ {\bf g}({\bf \tilde r}_\gamma, {\bf \tilde S}_\gamma) {\bm \cdot}
\partial/\partial {\bf \tilde q} \equiv \tilde {\mathcal B}_\gamma$
at $\alpha=-1/12$, $\beta=1/12$, and $\gamma=-\alpha=1/12$. Here
${\bf \Theta}\big({\bf J}^{-1}{\bf S}{\bf g}({\bf r},{\bf S}),\alpha
h^2 \big)=\exp[{\bf W}( {\bf Q}) \alpha h^2]$ is the matrix
representing three-dimensional rotation (i.e. $ {\bf \Theta}({\bf
Q},\tau) = {\bf I} \cos(Q \tau) + [1- \cos(Q \tau)] [ {\bf W}({\bf
Q}) {\bf W}({\bf Q})/Q^2 + {\bf I}] + \sin(Q \tau) {\bf W}({\bf
Q})/Q$ with ${\bf I}$ being the unit matrix) around vector ${\bf
Q}={\bf J}^{-1}{\bf S}{\bf g}$ on angle $Q \alpha h^2$, and $\{{\bf
\tilde r}_\gamma, {\bf \tilde S}_\gamma\} = \mathfrak{T}_{\gamma, 0}
\{{\bf \tilde r}, {\bf \tilde S}\}$ are the auxiliary position and
attitude matrix.

From the aforesaid, we have for the one-step propagation in
$\bm{\tilde \rho}$-space that $\bm{\tilde \rho}(t+h) = {\rm
e}^{\tilde L h} \bm{\tilde \rho}(t)= {\rm e}^{\tilde {\mathcal
B}_\gamma \frac{h}{2}} {\rm e}^{{\tilde \mathcal A} h} {\rm
e}^{\tilde {\mathcal B}_\gamma \frac{h}{2}} \bm{\tilde
\rho}(t)+{\mathcal O}(h^5)$. In ${\bm \rho}$-space the solution can
be reproduced by applying the inverse transformation
$\mathfrak{T}_{\alpha, \beta}^{-1}$ as $\bm{\rho}(t+h)={\rm e}^{L h}
\bm{\rho}(t) = \mathfrak{T}_{\alpha, \beta}^{-1} \bm{\tilde
\rho}(t+h)$. This leads to the resulting propagation of ${\bm \rho}$
in the form
\begin{equation}
{\rm e}^{L h} = \mathfrak{T}_{\alpha,\beta}^{-1} {\rm e}^{\tilde
{\mathcal B}_\gamma \frac{h}{2}} {\rm e}^{\tilde {\mathcal A} h}
{\rm e}^{\tilde {\mathcal B}_\gamma \frac{h}{2}}
\mathfrak{T}_{\alpha,\beta} + {\mathcal O}(h^5) ,
\end{equation}
where $\alpha=-1/12$, $\beta=1/12$, and $\gamma=1/12$. The operator
$\mathfrak{T}_{\alpha, \beta}$ transforms a phase space point
$\bm{\rho}$ to the set $\bm{\tilde \rho}=\mathfrak{T}_{\alpha,\beta}
\bm{\rho} \equiv \{{\bf \tilde r}, {\bf \tilde p}, {\bf \tilde S},
{\bf \tilde q}\}$ of time-step dependent pseudo-variables, where
\begin{eqnarray}
{\bf \tilde r} &=& {\bf r} + \alpha m^{-1} {\bf f}({\bf r},{\bf S})
h^2, \ \ \ \ \ \ \ \ \ \ \,
{\bf \tilde p} = {\bf p} + \beta {\bf \dot f}({\bm \rho}) h^2 ,
\nonumber \\ [-5pt] \\ [-5pt]
{\bf \tilde S} &=& {\bf \Theta}\big({\bf J}^{-1}{\bf S}
{\bf g}({\bf r},{\bf S}), \alpha  h^2 \big) {\bf S}, \ \ \ \
{\bf \tilde q} = {\bf q} + \beta {\bf \dot g}({\bm \rho}) h^2 .
\ \ \ \ \ \nonumber
\end{eqnarray}
The action of the exponential operators ${\rm e}^{\tilde {\mathcal
A} \tau}$ and ${\rm e}^{\tilde {\mathcal B}_\gamma \tau}$ can be
given analytically as
\begin{eqnarray}
{\rm e}^{\tilde {\mathcal A} \tau} \bm{\tilde \rho} &=& \big\{ {\bf
\tilde r} + m^{-1} {\bf \tilde p} \tau, {\bf \tilde p}, {\bf
\Xi}({\bf \tilde q}, \tau) {\bf \tilde S}, {\bf \tilde q} \big\} ,
\nonumber \\ [-6pt] \\ [-4pt]
{\rm e}^{\tilde {\mathcal B}_\gamma \tau} \bm{\tilde \rho} &=&
\big\{ {\bf \tilde r}, {\bf \tilde p} + {\bf f}({\bf \tilde
r}_\gamma,{\bf \tilde S}_\gamma) \tau, {\bf \tilde S}, {\bf
\tilde q} + {\bf g}({\bf \tilde r}_\gamma,{\bf \tilde S}_\gamma)
\tau \big\} . \ \ \ \ \nonumber
\end{eqnarray}
Expressions (5) are similar to Eq.~(2), since besides the formal
replacement of ${\bm \rho}$ by $\bm{ \tilde \rho}$ the only
difference between $(A,B)$ and $({\tilde \mathcal A},{\tilde
\mathcal B})$ lies in the modification of the force ${\bf f}({\bf
\tilde r}_\gamma,{\bf \tilde S}_\gamma)$ and torque ${\bf g}({\bf
\tilde r}_\gamma,{\bf \tilde S}_\gamma)$. Apart from the calculation
of their basic values ${\bf f}({\bf \tilde r},{\bf \tilde S})$ and
${\bf g}({\bf \tilde r},{\bf \tilde S})$, the modification requires
(for $\gamma \ne 0$) one extra force-torque evaluation at the
auxiliary positional ${\bf \tilde r}_\gamma = {\bf \tilde r} +
\gamma m^{-1} {\bf f}({\bf \tilde r}, {\bf \tilde S}) h^2$ and
orientational ${\bf \tilde S}_\gamma = {\bf \Theta}\big( {\bf
J}^{-1}{\bf \tilde S}{\bf g}({\bf \tilde r},{\bf \tilde S}), \gamma
h^2 \big) {\bf \tilde S}$ coordinates. This increases the number of
force-torque calculations in ${\rm e}^{\tilde {\mathcal B}_\gamma
\tau}$ from $n=1$ (at $\gamma=0$) to $n=2$ (at $\gamma \ne 0$), but
the order of precision of the processed splitting propagation grows
from $K=2$ (at $\alpha=\beta=\gamma=0$ when it reduces to the
genuine Verlet signature) to $K=4$ (at $-\alpha=\beta=\gamma=1/12$).

Because of $\mathfrak{T}^{-1}_{\alpha,\beta} \mathfrak{T}_{\alpha,
\beta}=1$, the solution to the equations of motion can now be cast
for any $t$ as $\bm{\rho}(t) = \mathfrak{T}^{-1}_{\alpha,\beta} [
{\rm e}^{\breve B_\gamma \frac{h}{2}} {\rm e}^{A h} {\rm e}^{\breve
B_\gamma \frac{h}{2}} ]^k \mathfrak{T}_{\alpha,\beta} \bm{\rho}(0) +
{\mathcal O}(h^4)$. Then the processing transformation
$\mathfrak{T}_{\alpha,\beta}$ can be performed only once on the very
beginning, while the inverse transformation $\mathfrak{T}_{\alpha,
\beta}^{-1}$ only once at the end of the considered time interval
$[0,t]$. In view of this, the step by step integration can be
interpreted as the time propagation of pseudo-variables $\bm{\tilde
\rho}$ by the kernel splitting ${\rm e}^{\breve B_\gamma \frac
{h}{2}} {\rm e}^{A h} {\rm e}^{\breve B_\gamma \frac{h}{2}}$ in the
transformed phase space. The real phase coordinates $\bm{\rho}$ are
not involved explicitly into the consecutive updating process. They
can be reproduced from $\bm{\tilde \rho}$ whenever it is necessary
(for example, when the measurement is desired) using the inverse
transformation $\bm{\rho}=\mathfrak{T}_{\alpha,\beta}^{-1} \bm{
\tilde \rho}$. This transformation reads (cf. to Eq.~(4))
\begin{eqnarray}
{\bf r} &=& {\bf \tilde r} - \alpha m^{-1} {\bf f}({\bf \tilde r},
{\bf \tilde S}) h^2, \ \ \ \ \ \ \ \ \ \ \ \ \ \ \,
{\bf p} = {\bf \tilde p} - \beta {\bf \dot f}(\bm{\tilde \rho})
h^2 , \nonumber \\ [-4pt] \\ [-6pt]
{\bf S} &=& {\bf \Theta}\big(\!- {\bf J}^{-1}{\bf \tilde S}
{\bf g}({\bf \tilde r},{\bf \tilde S}), \alpha h^2 \big)
{\bf \tilde S}, \ \ \ \
{\bf q} = {\bf \tilde q} - \beta {\bf \dot g}(\bm{\tilde \rho})
h^2 , \ \ \ \ \
\nonumber
\end{eqnarray}
where the higher-order terms ${\mathcal O}(h^4)$ have been neglected
since they are not accumulated in $\bm{\rho}(t)$.

The next crucial point concerns the evaluation of time derivatives
${\bf \dot f}(\bm{\tilde \rho})$ and ${\bf \dot g}(\bm{\tilde
\rho})$ which arise in Eq.~(6). It is obvious that their direct
evaluation should be obviated since this results in complicated
gradient terms. Fortunately, the derivatives can be evaluated at a
given $t$ in a quite efficient way by the symmetric interpolation
$\{ {\bf \dot f}, {\bf \dot g} \}(\bm{\tilde \rho})= [\{ {\bf \tilde
f}, {\bf \tilde g} \}(t+h) - \{ {\bf \tilde f}, {\bf \tilde g}
\}(t-h)]/(2h)+{\mathcal O}(h^2)$, where $\{ {\bf \tilde f}, {\bf
\tilde g}\}(t \pm h)= \{ {\bf f}, {\bf g}\}\big({\bf \tilde r}(t \pm
h),{\bf \tilde S}(t \pm h)\big)$. Such an interpolation is indeed
realizable because the pseudo-variables $\bm{\tilde \rho}(t\pm h)$
are determined step by step in the course of the kernel propagation
independently of $\bm{\rho}(t)$. Then the real variables
$\bm{\rho}(t)$ can be reproduced from $\bm{\tilde \rho}(t)$ with a
one-step retardation, when the pseudo-phase coordinates were already
propagated to $\bm{ \tilde \rho}(t+h)$. This avoids the calculation
of extra forces and torques during the interpolation and involves
only those which already were evaluated within the kernel
propagation. The time derivatives ${\bf \dot f}({\bm \rho})$ and
${\bf \dot g}({\bm \rho})$ in Eq.~(4) can be evaluated as $\{ {\bf
\dot f}, {\bf \dot g} \}({\bm \rho}) = [\{ {\bf f}, {\bf g} \}
(\frac{h}{2})- \{ {\bf f}, {\bf g} \}(-\frac{h}{2})]/h+{\mathcal
O}(h^2)$, where $\{ {\bf f}, {\bf g}\}(\pm \frac{h}{2})= \{ {\bf f},
{\bf g}\}\big({\bf r}(\pm \frac{h}{2}),{\bf S}(\pm \frac{h}{2})
\big)$ with ${\bf r}(\pm \frac{h}{2})={\bf r}(0) \pm m^{-1} {\bf
p}(0) \frac{h}{2}$ and ${\bf S}(\pm \frac{h}{2})={\bf \Theta}\big(
\pm {\bf J}^{-1}{\bf S}(0){\bf q}(0), \frac{h}{2}\big) {\bf S}(0)$.
This involves two extra forces and torques at $\pm h/2$ but
exclusively on the first step of the integration when starting from
an initial configuration $\bm{\rho}(0)$ and performing the direct
transformation $\mathfrak{T}_{\alpha,\beta}$.

We see therefore that the processed splitting (PS) algorithm derived
is truly of the fourth order and requires only $n=2$ force-torque
evaluations per time step. This overcomes the barrier $n=3$ inherent
in standard schemes. Moreover, the algorithm is time reversible
[because the exponential operators enter symmetrically into the
propagator (Eq.~(3))] and phase-area preserving [since simple shifts
and rotations (Eq.~(5)) do not change the volume]. In addition, the
algorithm is explicit (no iterations) and exactly conserves the
rigid molecular structure (because ${\bf \Xi}$ and ${\bf \Theta}$
are rotational matrices). The kernel splitting can also be made
symplectic, because it is based on the Verlet-like signature which
at $\gamma = 0$ conserves a nearby Hamiltonian \cite{Omejcp,Ramses,%
Ramsesi}). For a finite $\gamma \ne 0$, the potential operator can
be represented by $\tilde \mathcal B_\gamma = \tilde \mathcal B_0 +
\gamma [\tilde \mathcal B_0,[ \tilde \mathcal A,\tilde \mathcal
B_0]] h^2/2 + {\mathcal O}(h^4)$, where $[\tilde \mathcal B_0,[
\tilde \mathcal A,\tilde \mathcal B_0]]= (\tilde \mathcal
B_\varepsilon-\tilde \mathcal B_0)/\varepsilon + {\mathcal O}
(\varepsilon h^4)$ with $\varepsilon \ll 1$. Then the modified force
and torque in $\tilde \mathcal B_\gamma$  can be evaluated as ${\bf
f}({\bf \tilde r}_\gamma,{\bf \tilde S}_\gamma)= {\bf f}({\bf \tilde
r},{\bf \tilde S})+\gamma{\bf \Delta f}({\bf \tilde r},{\bf \tilde
S})$ and ${\bf g}({\bf \tilde r}_\gamma,{\bf \tilde S}_\gamma)={\bf
g}({\bf \tilde r},{\bf \tilde S})+\gamma{\bf \Delta g}({\bf \tilde
r},{\bf \tilde S})$, where the secondary fields are ${\bf \Delta
f}({\bf \tilde r},{\bf \tilde S}) = [{\bf f}({\bf \tilde
r}_\varepsilon,{\bf \tilde S}_\varepsilon) - {\bf f}({\bf \tilde
r},{\bf \tilde S})]/ \varepsilon + {\mathcal O}(\varepsilon h^4)$
and ${\bf \Delta g}\big({\bf \tilde r},{\bf \tilde S}) = [{\bf
g}({\bf \tilde r}_\varepsilon,{\bf \tilde S}_\varepsilon) - {\bf
g}({\bf \tilde r},{\bf \tilde S})]/\varepsilon + {\mathcal O}
(\varepsilon h^4)$. The parameter $\varepsilon$ is typically taken
to be of order $10^{-4}$ for double precision arithmetic to minimize
the effect of ${\mathcal O}(\varepsilon h^4)$-terms while avoiding
round-off truncations. The processing transformations (Eqs.~(4) and
(6)) need not be necessarily symplectic, since their effects are not
propagated ($\mathfrak{T}^{-1}_{\alpha,\beta}
\mathfrak{T}_{\alpha,\beta}=1$).

That is very surprising, within the PS method the number $n$ of
force-torque recalculations per time step can be reduced to $n=1$ at
all when a quasi-fourth order is requested. Note that the true
fourth order means that the deviations of the generated trajectories
$\bm{\rho}(t)$ from their exact counterparts are equal to ${\mathcal
O}(h^4) \sim C h^4$ at $t \gg h$. In MD simulations, this strong
requirement may not be so needed, because according to the Lyapunov
theorem \cite{Frenkel} the coefficient $C \sim {\rm e}^{\lambda t}$
grows ($\lambda > 0$) exponentially with increasing $t$. Then the
concept of the quasi-fourth order can be more useful. It implies
that the deviations apply not to individual variables of each
particle but rather to a collective function for which $C$ is
independent of $t$. In microcanonical simulations such a function
should be the total energy $E=\frac12 \sum_{i=1}^N ({{\bf p}_i}^2/m
+ {\bf \Omega}_i {\bf J} {\bf \Omega}_i) + \frac12 \sum_{i \ne
j;a,b}^{N;M} \varphi^{ab}(|{\bf r}_i-{\bf r}_j|, {\bf S}_i, {\bf
S}_j)$ of the system, where $\varphi^{ab}$ denotes the
intermolecular atom-atom potentials, and $M$ is the number of atoms
per molecule. Cumbersome analysis shows that $E$ can be conserved
with the fourth-order accuracy at $n=1$ by tuning the parameters of
the method to $\alpha=-1/24$, $\beta=1/12$, $\gamma=0$, and
$\eta=1/48$ (then ${\bm \rho}(t)$ and other quantities will not be
necessarily reproduced up to the fourth order). Here we should add a
new $\eta$-term when transforming (Eq.~(6)) angular momentum as
${\bf q} = {\bf \tilde q} - \beta {\bf \dot g}(\bm{\tilde \rho}) h^2
+ \eta {\bf \tilde S}^+[{\bf J}{\bf W}({\bf J}^{-1}{\bf \tilde
S}{\bf \tilde q}){\bf J}^{-1} - {\bf W}({\bf J}^{-1}{\bf \tilde
S}{\bf \tilde q}) - {\bf W}({\bf \tilde S}{\bf \tilde q}){\bf
J}^{-1}] {\bf \tilde S} {\bf g}({\bf \tilde r},{\bf \tilde S}) h^2$,
where ${\bf \tilde S}^+$ is the transposed matrix (and
correspondingly modify Eq.~(4)). This adding presents no difficulty
since ${\bf g}({\bf \tilde r},{\bf \tilde S})$ was already
calculated during the kernel splitting. For systems without periodic
boundary conditions, e.g. in celestial mechanics, the total angular
momentum is often also conserved. It will be kept with the
second-order accuracy by the quasi-fourth integrator ($n=1$). This
is in contrast to the genuine fourth-order algorithm ($n=2$) which
produces all quantities to within the ${\mathcal O}(h^4)$ precision.
Therefore, the former integrator may be less universally applicable
than the latter one. The PS algorithms will be referred to as PS1
($n=1$) and PS2 ($n=2$), respectively.

Further improvements are possible by splitting the atom-atom
potentials into short- and long-range parts. Then a multiple-time
stepping (MTS) technique \cite{Tuckerman} can be employed, where the
expensive long-range (weak) forces are sampled less frequently using
larger time steps, while the short-range (strong) interactions are
integrated more accurately inside the kernel propagator using
smaller steps. The MTS implementation within the PS method goes
beyond the scope of this paper and will be considered elsewhere.

\section{Numerical results}

We first present the proposed PS method (see Sec. II) in algorithmic
form to simplify its implementation in a numerical code. Thus,
starting from an initial configuration ${\bm \rho}(0) = \{ {\bf
r}(0), {\bf p}(0), {\bf S}(0), {\bf q}(0) \}$ at $t=0$ and
calculating the three forces ${\bf f}(0)$ and ${\bf f}({\pm h/2})$
as well as the three torques ${\bf g}(0)$ and ${\bf g}({\pm h/2})$
at the positions $\{ {\bf r}(0), {\bf S}(0) \}$ and $\{ {\bf r}(\pm
h/2), {\bf S}(h/2) \}$, respectively, where ${\bf r}(\pm h/2)={\bf
r}(0) \pm m^{-1} {\bf p}(0) h/2$ and ${\bf S}(\pm h/2)={\bf
\Theta}\big(\pm {\bf J}^{-1}{\bf S}(0){\bf q}(0), h/2\big) {\bf
S}(0)$, we make the direct processing transformation (Eq.~(4)) to
$\bm{\tilde \rho}(0) = \{ {\bf \tilde r}(0), {\bf \tilde p}(0), {\bf
\tilde S}(0), {\bf \tilde q}(0) \}$ as
\begin{eqnarray}
{\bf \tilde r}(0) &=& {\bf r}(0) + \alpha m^{-1} {\bf f}(0) h^2 ,
\nonumber \\
{\bf \tilde S}(0) &=& {\bf \Theta}\big( {\bf J}^{-1}{\bf S}(0)
{\bf g}(0), \alpha  h^2 \big) {\bf S}(0) ,
\nonumber \\ [-7pt] \\ [-7pt]
{\bf \tilde p}(0) &=& {\bf p}(0) + \beta \big(
{\bf f}(h/2)-{\bf f}(h/2)\big) h ,
\nonumber \\
{\bf \tilde q}(0) &=& {\bf q}(0) + \beta \big(
{\bf g}(h/2)-{\bf g}(h/2)\big) h .
\nonumber
\end{eqnarray}
Having $\bm{\tilde \rho}(0)$, we calculate the two initial forces
${\bf \tilde f}(0)$ and ${\bf \tilde f}_\varepsilon(0)$ as well as
the two initial torques ${\bf \tilde g}(0)$ and ${\bf \tilde
g}_\varepsilon(0)$ at the positions $\{ {\bf \tilde r}(0), {\bf
\tilde S}(0) \}$ and $\{ {\bf \tilde r}_\varepsilon(0), {\bf \tilde
S}_\varepsilon(0) \}$, respectively, where ${\bf \tilde
r}_\varepsilon(0) = {\bf \tilde r}(0) + \varepsilon m^{-1} {\bf
\tilde f}(0) h^2$ and ${\bf \tilde S}_\varepsilon(0) = {\bf
\Theta}\big( {\bf J}^{-1}{\bf \tilde S(0)}{\bf \tilde g}(0),
\varepsilon h^2 \big) {\bf \tilde S}(0)$. Note that the direct
transformation (Eq.~(7)) as well as the evaluation of the initial
forces and torques should be carried out only once at the very
beginning ($t=0$) of the integration.

Now we perform the single-step propagations of $\bm{\tilde \rho}$
from time $t$ to $t+h$ according to the kernel splitting (Eq.~(3))
as
\begin{eqnarray}
{\bf \tilde p}_{t+\frac{h}{2}} &=& {\bf \tilde p}(t) +
\Big[{\bf \tilde f}(t) + \frac{\gamma}{\varepsilon} \big(
{\bf \tilde f}_\varepsilon(t)-{\bf \tilde f}(t) \big) \Big]
\frac{h}{2} ,
\nonumber \\
{\bf \tilde q}_{t+\frac{h}{2}} &=& {\bf \tilde q}(t) +
\Big[{\bf \tilde g}(t) + \frac{\gamma}{\varepsilon} \big(
{\bf \tilde g}_\varepsilon(t)-{\bf \tilde g}(t) \big) \Big]
\frac{h}{2} ,
\nonumber \\
{\bf \tilde r}(t+h) &=& {\bf \tilde r}(t) +
m^{-1} {\bf \tilde p}_{t+\frac{h}{2}} h ,
\nonumber \\ [-6pt] \\ [-6pt]
{\bf \tilde S}(t+h) &=& {\bf \Xi}\big(
{\bf \tilde q}_{t+\frac{h}{2}},h\big) {\bf \tilde S}(t) ,
\nonumber \\
{\bf \tilde p}(t+h) \!&=&\! {\bf \tilde p}_{t+\frac{h}{2}} +
\Big[{\bf \tilde f}(t\!+\!h) + \frac{\gamma}{\varepsilon} \big(
{\bf \tilde f}_\varepsilon(t\!+\!h) - {\bf \tilde f}(t\!+\!h) \big)
\Big] \frac{h}{2} ,
\nonumber \\
{\bf \tilde q}(t+h) \!&=&\! {\bf \tilde q}_{t+\frac{h}{2}} +
\Big[{\bf \tilde g}(t\!+\!h) + \frac{\gamma}{\varepsilon} \big(
{\bf \tilde g}_\varepsilon(t\!+\!h) - {\bf \tilde g}(t\!+\!h) \big)
\Big] \frac{h}{2} ,
\nonumber
\end{eqnarray}
where ${\bf \tilde p}_{t+\frac{h}{2}}$ and ${\bf \tilde q}_{t+
\frac{h}{2}}$ are the intermediate values, and the two new forces
${\bf \tilde f}(t+h)$ and ${\bf \tilde f}_\varepsilon(t+h)$ as well
as the two new torques ${\bf \tilde g}(t+h)$ and ${\bf \tilde
g}_\varepsilon(t+h)$ should be calculated at the new positions $\{
{\bf \tilde r}(t+h), {\bf \tilde S}(t+h) \}$ and $\{ {\bf \tilde
r}_\varepsilon(t+h), {\bf \tilde S}_\varepsilon(t+h) \}$,
respectively, with ${\bf \tilde r}_\varepsilon(t+h) = {\bf \tilde
r}(t+h) + \varepsilon m^{-1} {\bf \tilde f}(t+h) h^2$ and ${\bf
\tilde S}_\varepsilon(t+h) = {\bf \Theta}\big( {\bf J}^{-1}{\bf
\tilde S}(t+h){\bf \tilde g}(t+h), \varepsilon  h^2 \big) {\bf
\tilde S}(t+h)$ before the evaluation of ${\bf \tilde p}(t+h)$ and
${\bf \tilde q}(t+h)$. Saving the forces ${\bf \tilde f}(t+h)$ and
${\bf \tilde f}_\varepsilon(t+h)$ as well as the torques ${\bf
\tilde g}(t+h)$ and ${\bf \tilde g}_\varepsilon(t+h)$, we repeat
Eq.~(8) (with formal replacing $t$ by $t+h$ in it) to propagate
$\bm{\tilde \rho}$ from time $t+h$ to $t+2h$. In such a way, step by
step we can recycle Eq.~(8) arbitrarily number $k \ge 1$ of times
and obtain the value of $\bm{\tilde \rho}(t)$ for any $t=k h$. Each
recycle will require the recalculation of only two ($n=2$) new
forces and torques.

When at least two recycles of Eq.~(8) are done already, we will have
the three consecutive values $\bm{\tilde \rho}(t-h)$, $\bm{\tilde
\rho}(t)$, and $\bm{\tilde \rho}(t+h)$ for some $t=k h$. The forces
${\bf \tilde f}(t)$ and ${\bf \tilde f}(t \pm h)$ as well as the
torques ${\bf \tilde g}(t)$ and ${\bf \tilde g}(t \pm h)$ will also
be already known because of the kernel propagations. Then we can
make the inverse processing transformation (Eq.~(6)) of $\bm{\tilde
\rho}(t)$ to the genuine value ${\bm \rho}(t)$ at a current $t$
according to
\begin{eqnarray}
{\bf r}(t) &=& {\bf \tilde r}(t) - \alpha m^{-1} {\bf \tilde f}(t)
h^2 ,
\nonumber \\
{\bf S}(t) &=& {\bf \Theta}\big(\!- {\bf J}^{-1}{\bf \tilde S}(t)
{\bf \tilde g}(t), \alpha  h^2 \big) {\bf \tilde S}(t) ,
\nonumber \\ [-7pt] \\ [-7pt]
{\bf p}(t) &=& {\bf \tilde p}(t) - \beta \big(
{\bf \tilde f}(t+h)-{\bf \tilde f}(t-h)\big) h/2 ,
\nonumber \\
{\bf q}(t) &=& {\bf \tilde q}(t) - \beta \big(
{\bf \tilde g}(t+h)-{\bf \tilde g}(t-h)\big) h/2 ,
\nonumber
\end{eqnarray}
and calculate at this point all necessary observable quantities
(such as the total energy, etc.). This completes the PS2 algorithm
($n=2$), where $\alpha=-1/12$, $\beta=1/12$, and $\gamma=1/12$. The
PS1 integrator ($n=1$) follows at $\alpha=-1/24$, $\beta=1/12$,
$\gamma=0$, and $\eta=1/48$ (here the evaluation of the modified
force ${\bf \tilde f}_\varepsilon$ and torque ${\bf \tilde
g}_\varepsilon$ should be omitted in Eq.~(8) since $\gamma=0$, while
the inclusion of the $\eta$-term in Eqs.~(7) and (9) is trivial).

For testing of the algorithms we applied the TIP4P model ($M=4$) of
water \cite{Jorgensen} with $N=512$ molecules. The MD simulations
were carried in the microcanonical ($NVE$) ensemble at a density of
$N/V=$ 1 g/cm$^3$ and a temperature of 292 K. The Ewald summation
\cite{OEwald} was exploited to handle long-range Coulombic atom
interactions. The accuracy of the simulations was measured by
calculating the ratio ${\mathcal R}$ of the fluctuations of the
total energy $E$ to the fluctuations of its potential part
\cite{Omejcp}. The computational costs $\Upsilon$ were estimated in
terms of the number of force-torque evaluations in a given time
interval, taken to be $\Lambda=1$ ps, so that $\Upsilon=n
\Lambda/h$. The equations of motion were solved at several sizes of
the time step ranging from $h = 0.5$ fs to 5 fs. In total $k=t/h =
10^5$ steps were used for each algorithm and each step size.

The costs $\Upsilon$ versus precisions ${\mathcal R}$ of the
integration obtained within the two proposed PS algorithms ($K=4$)
at the end of the simulations are plotted in Fig.~1 by the curves
marked as PS1 ($n=1$) and PS2 ($n=2$), respectively. The results
corresponding to the Verlet-type (VT) algorithm ($K=2$ and $n=1$),
its optimized (VO) version ($K=2$ and $n=2$), the Forest-Ruth (FR)
scheme ($K=4$ and $n=3$), as well as the gradient-like (GL)
algorithm ($K=4$ and $n=3$) (these integrators are described in
Ref.~\cite{Omejcp,Ramsesi}) were also included for the purpose of
comparison. It has been established that other known rigid-body
integrators \cite{Kol,Dullw,Matubayasi,Miller,Kamberaj,Ryckaert}
($K=2$ and $n=1$) behave similarly to the VT algorithm. Higher-order
schemes \cite{Omejcp,Blanesa} with $K \ge 4$ and $n \ge 4$ are less
efficient in MD simulations because of the large numbers of costly
force-torque recalculations. The processed fourth-order algorithm by
Blanes and Casas (BC) {\em et al.} \cite{Blanes,Casas} with $K=4$
and $n=1+\nu=3$ (where the kernel and processor are defined
according to Eqs.~(20) and (21) of Ref.~\cite{Blanes}) was adapted
to rigid-body motion and considered too.

\begin{figure}
\includegraphics[width=84mm]{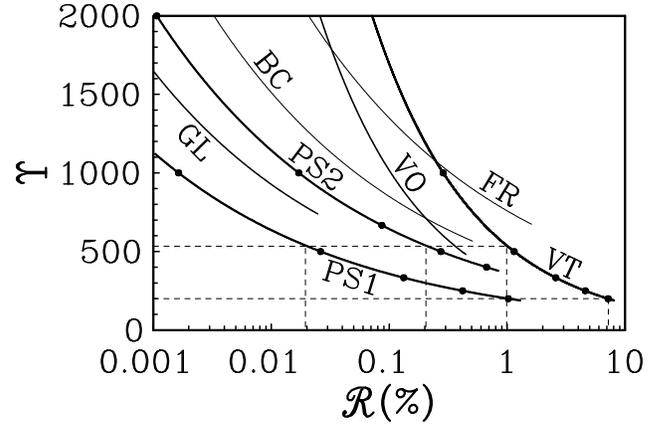}
\caption{The cost versus relative error for different algorithms in
MD simulations of water. The circles correspond to the time steps
(left to right) $h=1,2,3,4$, and 5 fs. The dashed lines represent
the most characteristic levels.}
\end{figure}

As can be seen from Fig.~1, with decreasing $\Upsilon$ (rising $h$)
each curve terminates at some point where the simulations begin to
exhibit a drift in ${\mathcal R}$. This happens around $h \sim 5$ fs
(larger $h$ can be used within the MTS). At the minimally possible
costs $\Upsilon \sim 200$, the VT integrator can provide only a
crude energy conservation ${\mathcal R} \sim 7 \%$. This level of
errors is too large and generally unacceptable in MD simulations. It
should be reduced at least to ${\mathcal R} \sim 1 \%$, arguably the
upper limit of allowable error for which the dynamics can be
simulated adequately. The proposed PS1 algorithm just satisfies this
criteria even at $\Upsilon \sim 200$. On the other hand, the level
${\mathcal R} \sim 1 \%$ can be achieved by the VT integrator by
increasing the load to $\Upsilon \sim 550$, i.e. in a factor of
2.75. Thus the PS1 algorithm may spend considerably smaller CPU time
at a given precision. The PS2 algorithm is also superior to the VT
scheme. For more accurate (${\mathcal R} < 1 \%$) simulations, the
relative efficient of the PS algorithms ($K=4$) with respect to the
VT scheme ($K=2$) rises further (because ${\mathcal R} \sim h^K$)
and reaches a factor of 5 at ${\mathcal R} \sim 0.1 \%$. At the same
time, for $\Upsilon \sim 550$ the PS2 and PS1 algorithms are able to
lower the numerical errors from the value ${\mathcal R} \sim 1\%$
inherent in the VT integrator to the levels ${\mathcal R} \sim
0.2\%$ and 0.02\%, respectively, i.e. up in 50 times! The VO
integrator is clearly inferior to the PS algorithms, although it can
be better than the VT signature. The BC scheme is superior to the VO
integrator but worse than the PS algorithms. The FR scheme leads to
the worst efficiency. The GL algorithm can be used only at $\Upsilon
> 750$, i.e. when a very high accuracy (${\mathcal R} \lesssim 0.02
\%$) is required. Then it appears to be more efficient than the PS2
integrator. However, the PS1 algorithm is the best in the whole
$\Upsilon$-region.

Samples of the relative fluctuations ${\mathcal R}(t)$ and
normalized deviations $\delta E(t)=(E(t)-E(0))/E(0)$ of the
instantaneous total energy $E(t)$ are shown in subsets (a) and (b)
of Fig.~2, respectively, versus the length $t/h$ of the simulations
performed at a typical step $h=4$ fs using different integrators. We
can observe in Fig.~2(a) that the functions ${\mathcal R}(t)$ are
flat with no drift on the entire time domain. The PS algorithms thus
apart from their high efficiency, exhibit also excellent stability
properties. As is illustrated in Fig.~2(b) for the PS1 method, the
total energy $E(t)$ continues to keep near its initial value $E(0)$
even after an extremely long period of time with $k=10^6$ steps. The
magnitude of the deviations $\delta E(t)$ is quite small and does
not exceed a level of 0.01\%, making the energy conservation almost
exact.

\begin{figure}
\includegraphics[width=86mm]{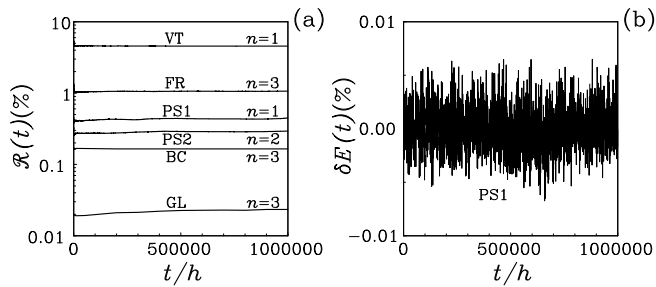}
\caption{The fluctuations (a) and deviations (b) of the total energy
versus the length of the MD simulations carried out at $h=4$ fs
using different algorithms.}
\end{figure}

\section{Conclusion}

In this paper we have proposed a novel method for the integration of
motion in rigid-body MD simulations. It combines standard splitting
techniques with special phase-space processing transformations.
Comparison with the well-recognized previous schemes has
demonstrated that the new method allows to significantly improve the
efficiency of the integration with no extra computational costs. The
algorithms obtained are easy in implementation and can readily be
incorporated into existing MD codes. They can also be applied to
hybrid Monte-Carlo, MD simulations of simple fluids and to other
fields mentioned in the introduction as well as be extended to more
complicated systems with flexible molecules.

\section*{ACKNOWLEDGMENT}

The author acknowledges support by the Fonds zur F\"orderung der
Wissenschaftlichen Forschung under the Project No. P18592-TPH.


\begin{thebibliography}{99}

\bibitem{Allen}
 M. P. Allen and D. J. Tildesley, {\em Computer Simulation of
 Liquids} (Clarendon, Oxford, 1987).

\bibitem{Rapaport}
 D. C. Rapaport, {\em The Art of Molecular Dynamics Simulation}
 (Cambridge University Press, Cambridge, 1995).

\bibitem{Frenkel}
 D. Frenkel and B. Smit, {\em Understanding Molecular Simulation:
 from Algorithms to Applications} (Academic Press, New York, 1996).

\bibitem{Essiz}
 S. Essiz and R. D. Coalson,  J. Chem. Phys. {\bf 124}, 144116 (2006).

\bibitem{Celledoni}
 E. Celledoni and N. S\"afstr\"om,
 J. Phys. A, Math. Gen. {\bf 39}, 5463 (2006).
% Efficient time-symmetric simulation of torqued rigid bodies
% using Jacobi elliptic functions

\bibitem{Chin}
 S. A. Chin, Phys. Rev. E {\bf 75}, 036701 (2007).

\bibitem{Reichs}
 S. Reich, Fields Inst. Commun. {\bf 10}, 181 (1996).

\bibitem{Kol}
 A. Kol, B. B. Laird, and B. J. Leimkuhler,
 J. Chem. Phys. {\bf 107}, 2580 (1997).
% A symplectic method for rigid-body molecular simulation
% (RSHAKE)

\bibitem{Dullw}
 A. Dullweber, B. Leimkuhler, and R. McLachlan,
 J. Chem. Phys. {\bf 107}, 5840 (1997).
% Symplectic splitting methods for rigid body molecular dynamics
% (angular momentum in laboratory frame)

\bibitem{Matubayasi}
 N. Matubayasi and M. Nakahara,
 J. Chem. Phys. {\bf 110}, 3291 (1999).
% Reversible molecular dynamics for rigid bodies and hybrid Monte Carlo
% (principal angular velocity rotational)

\bibitem{Miller}
 T. F. Miller III, M. Eleftheriou, P. Pattnaik, A. Ndirango,
 D. Newns, and  G. J. Martyna,
 J. Chem. Phys. {\bf 116}, 8649 (2002).
% Symplectic quaternion scheme for biophysical molecular dynamics

\bibitem{Kamberaj}
 H. Kamberaj, R. J. Low, M. P. Neal, J. Chem. Phys. {\bf 122}, 224114 (2005).
% Time reversible and symplectic integrators for molecular dynamics
% simulations of rigid molecules
% (incorrect implementation of the triplet fourth-order rotational algorithm)

%\bibitem{Okumura}
% H. Okumura, S. G. Itoh, and Y. Okamoto, J. Chem. Phys. {\bf 126}, 084103 (2007).
%% Explicit symplectic integrators of molecular dynamics algorithms for rigid-body
%% molecules in the canonical, isobaric-isothermal, and related ensembles

\bibitem{Omfcpcn}
 I. P. Omelyan, I. M. Mryglod, and R. Folk, Comput. Phys. Commun.
 {\bf 151}, 272 (2003).

\bibitem{Omepre}
 I. P. Omelyan, Phys. Rev. E {\bf 74}, 036703 (2006).

\bibitem{Omejcp}
 I. P. Omelyan, J. Chem. Phys. {\bf 127}, 044102 (2007).

\bibitem{Blanes}
 S. Blanes, F. Casas, and J. Ros,
 SIAM (Soc. Ind. Appl. Math.) J. Sci. Stat. Comput. {\bf 21}, 711 (1999).
% Symplectic integration with processing: a general study

\bibitem{Skeel}
 R. D. Skeel, G. Zhang, and T. Schlick, SIAM (Soc. Ind. Appl.
 Math.) J. Sci. Stat. Comput. {\bf 18}, 203 (1997).

\bibitem{Marcos}
 M. A. L\'opez-Marcos, J. M. Sanz-Serna, and R. D. Skeel,
 SIAM (Soc. Ind. Appl. Math.) J. Sci. Stat. Comput. {\bf 18}, 223 (1997).

\bibitem{Casas}
 S. Blanes, F. Casas, and A. Murua,
 SIAM (Soc. Ind. Appl. Math.) J. Sci. Stat. Comput. {\bf 42}, 531 (2004);
% On the numerical integration of ordinary differential equations by processed methods
{\bf 27}, 1817 (2006).
% Composition methods for differential equations with processing

\bibitem{Ramses}
 R. van Zon and J. Schofield, Phys. Rev. E {\bf 75}, 056701 (2007).
% Symplectic algorithms for simulations of rigid-body systems using
% the exact solution of free motion

\bibitem{Ramsesi}
 R. van Zon, I. P. Omelyan, and J. Schofield,
 J. Chem. Phys. {\bf 128}, 136102 (2008).

\bibitem{Tuckerman}
 M. E. Tuckerman, B. J. Berne, and G. J. Martyna,
 J. Chem. Phys. {\bf 97}, 1990 (1992).

\bibitem{Jorgensen}
 W. L. Jorgensen, J. Chandrasekhar, J. D. Madura, R. W. Impey, and M. L. Klein,
 J. Chem. Phys. {\bf 79}, 926 (1983).

\bibitem{OEwald}
 I. P. Omelyan, Comput. Phys. Commun. {\bf 107}, 113 (1997).

\bibitem{Ryckaert}
 J. P. Ryckaert, G. Ciccotti, and H. J. C. Berendsen,
 J. Comput. Phys. {\bf 23}, 327 (1977).

\bibitem{Blanesa}
 S. Blanes, and F. Casas, J. Phys. A, Math. Gen. {\bf 39}, 5405 (2006).

\end{thebibliography}
\end{document}